\documentstyle[12pt,aaspp4]{article}
\def\ie{i.e.\ }
\def\eg{e.g.,\ }
\def\etal{et~al.\ }
\def\ltsima{$\; \buildrel < \over \sim \;$}
\def\simlt{\lower.5ex\hbox{\ltsima}}
\def\gtsima{$\; \buildrel > \over \sim \;$}
\def\simgt{\lower.5ex\hbox{\gtsima}}
\def\ha{H$\alpha$\ }
\def\kms{km~s$^{-1}$}

\def\fp{Fabry-P\'erot}
\def\degns{\ifmmode^\circ\else$^\circ$\fi}
\def\deg{\ifmmode^\circ\else$^\circ$\fi\ }
\journalid{}{}
\articleid{}{}
\lefthead{Mihos \& Bothun}
\righthead{Velocity Mapping of ULIRGs}
\slugcomment{to appear in the {\it Astrophysical Journal}}
 
\begin{document}
 
\title{\ha Velocity Mapping of Ultraluminous Infrared Galaxies}
 
\author{J. Christopher Mihos\altaffilmark{1,2,3}}
\affil{Department of Physics and Astronomy\break
         Johns Hopkins University, Baltimore, MD 21218\break
         hos@burro.astr.cwru.edu}

\and

\author{Gregory D. Bothun}
\affil{Department of Physics\break
         University of Oregon, Eugene, OR 97403\break
         nuts@moo.uoregon.edu}
 
\altaffiltext{1}{Hubble Fellow}
\altaffiltext{2}{Visiting Astronomer, Cerro Tololo Inter-American Observatory.
CTIO is operated by AURA, Inc. under contracts to the National Science
Foundation.}
\altaffiltext{3}{Current Address: Department of Astronomy, Case Western
Reserve University, 10900 Euclid Ave, Cleveland, OH 44106.}

\begin{abstract}

We use imaging \fp\ observations to explore the dynamical conditions in 
four southern ultraluminous ($L_{IR}$ \gtsima $10^{12}$ L$_{\sun}$) infrared 
galaxies. These galaxies all show morphological and kinematic features 
indicative of a major merger, but span a wide range of inferred dynamical 
ages: the youngest system is just past the first encounter, while the oldest
has already merged into a single object. This diversity in dynamical states
indicates that ultraluminous activity is {\it not} confined solely to 
late-stage mergers. The \ha
emission is more spatially concentrated in the later stage mergers, in
agreement with models which show that mergers drive gas inwards towards the
nucleus, but high luminosities are achieved even in young interactions
where the \ha is quite extended. Our data indicate that physical details
other than the dynamical phase -- such as the internal structure or gas content
of the galaxies -- can play a strong role in determining the luminosity 
evolution of merging galaxies.  We also find evidence for massive star-forming 
complexes at large radius in the tidal debris, reaffirming the notion that 
some dwarf galaxies may be spawned during the merger of more massive
disk galaxies.

\end{abstract}
 
\keywords{galaxies:active, galaxies:interactions, galaxies:{kinematics and 
dynamics}, galaxies:starburst, infrared:galaxies} 
 
\section{Introduction}

With bolometric luminosities rivaling those of quasars, the so-called
``ultraluminous infrared galaxies'' (ULIRGs) are the most luminous
objects in the local universe.   The basic properties of these galaxies
have been reviewed by Sanders and Mirabel (1996) and their most
outstanding characteristics can be summarized as follows: 1) 90\% of
their bolometric luminosity is emitted in the wavelength range 
12-100$\mu$; 2) with one exception, all have very high CO luminosities 
(see Solomon \etal 1997); 3) when imaged deep enough, all show signs
of strong interactions or mergers (see Melnick and Mirabel 1990,
Clements and Baker 1996;
Clements etal 1996a).  The phenomenological picture which emerges 
is that ULIRGS are the manifestation of the collision and subsequent 
merger of two disk systems which are rich in molecular gas.  
Support for this picture comes from the observed molecular gas masses of
ULIRGS which are consistent with what has been measured for gas-rich
luminous disk galaxies  (Solomon \etal 1997), as well as the refutation
of the assertion by Leech \etal (1994) that some ULIRGS were not
in strongly interacting systems.   Better imaging by Clements and
Baker (1996), Clements etal (1996ab) and Lucas etal (1997) has shown
that virtually all ULIRGS show morphological signatures of a major merger.

Behind the phenomenology, however, lies the physics which produces the
very high luminosities.  While the trigger for the effect may now be
clear, it remains controversial if starbursts ( Joseph \& Wright 1985)
or AGN activity (Sanders \etal 1988) are the main power source of the 
luminosity.  Both have been invoked and there is evidence to support
either case.  For instance, in some ULIRGs, emission line diagnostics 
indicate the presence of buried AGN in their nuclei, while other systems 
show no such evidence (Goldader \etal 1995), even in the mid-IR where the 
dust obscuration is less troublesome (Lutz \etal 1996). While the 
infrared spectra of ULIRGs are difficult to fit with a pure starburst model 
(Goldader \etal 1995), it remains unclear whether this is due to the 
presence of an AGN or simply a consequence of extreme optical depths in 
ULIRGs (Lutz \etal 1996).  Recent ASCA observations by Ogasaka \etal (1997)
reveal that at least one ULIRG has a 2-10 keV luminosity which is similar
to that observed in QSOs (\eg 10$^{44-45}$ ergs s$^{-1}$).  On the
other hand, VLA observations by Crawford etal (1996) shows that the
radio continuum luminosity of ULIRGs is consistent with the starburst driven
correlation between FIR and radio luminosity established earlier by
Helou \etal (1985).

In reality, both starburst and AGN activity may be occurring concurrently, with 
the detailed energy balance determined by a variety of factors and there
may well be some evolutionary connection between these two states.   Clearly,
the very high inferred densities of molecular gas (up to 10$^7$
cm$^{-3}$, Solomon \etal 1992) can generate huge starbusts.  Moreover, 
at these densities the disk is extremely viscous (see Bryant and
Scoville 1996). which  greatly facilitates the viscous transport of material 
onto a central engine, if one exists.  Hence there maybe valid physical
reasons to view ULIRGs as systems that are initially powered by starbursts
and then later powered by accretion onto a central engine and in that way
ULIRGs affirm the starburst-AGN connection.  Support for this hypothesis
also comes from the work of Chang \etal (1987) who show that the dynamical
lifetime of dust and other obscuring material which is in near a luminous
AGN is less than the radiating lifetime of the AGN.  Hence, if AGN
activity is the initial power source of the ULIRGs some fraction of them
should evolve to become unobscured QSOs but still show signs of 
merger activity.  One example of this evolution might be provided by the low
redshift QSO PKS 2349-014 (Bahcall \etal 1995).

To trigger either AGN or starburst activity, some mechanism must act to
transport gas down to the central regions of the ULIRGs. Galaxy interactions
and mergers seem the most likely mechanism to drive these central inflows of 
gas and this is consistent with the phenomenological picture painted 
above.  Dynamical modeling has demonstrated the efficacy of interactions at
driving inflows (Noguchi 1991; Barnes \& Hernquist 1991, 1996; Mihos \etal
1992; Hernquist \& Mihos 1995; Mihos \& Hernquist 1996). 
However, the details of this link between 
interactions and luminous activity are not well understood.  While most 
ULIRGs are found in closely interacting and/or merging systems, the converse 
is not true -- the great majority of interacting systems do not 
display ``ultraluminous'' levels of emission. 
Clearly a crucial element in driving central activity in ULIRGs must be the 
detailed dynamical conditions of the interacting galaxies. Models by Barnes 
\& Hernquist (1996) suggest that the onset and intensity of gas inflows is 
linked to the geometry and impact parameter of the galactic collision, while 
the dynamical role of central bulges in regulating inflows was demonstrated
by Mihos \& Hernquist (1996). Moreover, the build-up of molecular
gas to high densities may also act as an efficient damper of angular momentum
of additional gas which is introduced into the central regions thereby
effectively serving as a conduit to fuel an AGN.

A better understanding of the
conditions necessary for spawning ultraluminous activity can only come
through studies of the dynamics of ultraluminous infrared galaxies. 
Unfortunately, morphology alone gives only a weak handle on the dynamical 
state of the system. For example, while two galaxies may appear closely 
interacting, they may not exhibit a strong dynamical response, depending on 
the encounter geometry (prograde vs. retrograde encounters), the structural
properties of the galaxies, or the age of the interaction. These uncertainties
make it difficult to define the true dynamical state of many luminous infrared
galaxies. 
To better investigate the dynamical conditions in ULIRGs, we 
present \ha\ velocity 
maps of four southern ULIRG systems, taken with the Rutgers Imaging \fp\
spectrograph. Unlike slit spectra, imaging \fp\ data map the full two
velocity field of the galaxies, and the arcsecond spatial resolution achieved 
in the optical betters that of HI velocity maps by an order of magnitude.
Using the morphology and velocity field in conjunction with numerical 
modeling, it is possible to reconstruct much of the dynamical history of 
merging systems (\eg Borne, Balcells, \& Hoessel 1988; Stanford \& Balcells 
1991; Mihos \etal 1993; Hibbard \& Mihos 1995; Mihos \& Bothun 1997) and 
probe the dynamical conditions necessary for driving ``ultraluminous'' levels 
of activity.

\section{Observations}

Our sample of ultraluminous infrared galaxies is neither complete nor unbiased.
The four objects (Table 1) were chosen to have high far-IR luminosity
(L$_{IR}$ \gtsima 10$^{12}$ L$_{\sun}$) and extended, high surface brightness
tidal features. The morphological criteria were employed to maximize the
likelihood of detecting \ha in the tidal features, enabling us to map the
velocity field of the systems at large radii. However, selecting on the
presence of tidal debris probably biases our sample towards systems in
which at least one galaxy was a somewhat prograde disk, and also towards
systems which are relatively young. Tidal features are much harder to
detect in retrograde mergers or in very old remnants where the tidal debris
has become very diffuse.

The \ha\ velocity mapping described here was done using the Rutgers Imaging \fp\ 
on the CTIO 4-m telescope. The \fp\ etalons have a bandpass of 2.2\AA\ 
FWHM, giving an instrumental velocity dispersion of $\sim$ 50 \kms. To isolate 
the \ha\ transmission order, narrow band ($\sim$ 80\AA) \ha\ filters were 
used. Using the Tek 2k CCD, the pixel scale was 0.36\arcsec\ pixel$^{-1}$ and 
the circular field of view covered $\sim$ 1.5\arcmin\ in diameter. This
field of view was sufficient to encompass each system, with the exception of 
IRAS 19254-7245, whose long tidal tails extended off the edge of the field.

Observations of each object consist of a series of 10 minute exposures 
typically stepped 1\AA\ apart. The spectral range covered depends on the 
velocity width of the system; details for each object are given in Table 2.
We emphasize that observing time constraints made it impractical to 
extend spectral coverage in order to detect broad wings that might 
be associated with nuclear outflow or the presence of an AGN.  
Data reduction consisted of flat fielding, sky subtraction, 
and normalization of the individual images to a common transparency (see
Mihos \& Bothun 1997). Because the redshift range of the sample
and the radial wavelength gradient across the field of view,
night OH sky lines appear on the images as diffuse rings of emission.
These night sky rings are removed by masking out the object and creating
a map of the median intensity as a function of radius in each image, which
is then subtracted from the original. Transparency normalization was done by
assuming that stars in the field have a flat spectrum over the limited spectral
range scanned. Because of the small field of view, only a few
stars are available for photometry in each field, making accurate
normalization difficult. Typical corrections for transparency are 
less than 20\%, with errors of 5 to 10\%. Wavelength calibration was
achieved through a series of neon lamp exposures taken during the
night.

After the images are normalized, each pixel has a spectrum associated with
it which traces out the spectral shape of the \ha\ emission line at each point
in the galaxy. Binning the images 3x3 to improve signal, the spectra are fit
to a single Voigt profile using least squares minimization. From this fitting,
four parameters are extracted: the continuum level, the \ha\ line flux,
the central velocity of the line, and the gaussian velocity dispersion. 
Each pixel was fit automatically, after which the individual fits were examined
by eye while watching {\it Melrose Place}.
Fits to obvious noise were deleted, and bad or missing fits to good 
data were improved by giving the fitter a better initial estimate of the 
parameters.  In the final maps, typical errors on the derived velocity are 
$\pm$20 \kms, as long as the line profile was simple. However,
some regions proved difficult to fit to a single line profile, having
asymmetric lines, multiple components, or velocity widths greater
than the scanned spectral range. These regions are discussed in detail
in \S 3 below.

\section{Individual Objects}

\subsection{IRAS 14348-1447}

IRAS 14348-1447 (hereafter IR14348) is the most distant and infrared
luminous of the four galaxies observed (Table 1). The optical morphology
(Sanders \etal 1988; Melnick \& Mirabel 1990) is that of a double nucleus 
system, with a bright tidal tail extending to the north and a fainter
tail to the southwest. Near-IR imaging reveals 
two nuclei (Carico \etal 1990) with a separation of $\sim$
3.5\arcsec\ (5 kpc). The nuclei have a velocity difference of $\Delta v =
150$ \kms\ from optical emission lines, and the nuclear spectra show
indications of AGN activity, classified alternately as LINER (Veilleux
\etal 1995) or Seyfert 2 (Sanders \etal 1988; Nakajima \etal 1991). 
Single dish CO(1-2) measurements indicate the system contains $\sim
6 \times 10^{10}$ M$_{\sun}$ of molecular gas (Sanders, Scoville, \&
Soifer 1991). 

An examination of the fitted spectra showed no complex lines; all data
was reasonably well fit by a single component Voigt profile. However,
the spectral range scanned was not sufficient to sample the full velocity 
width of the Seyfert 2 line profile in the SW nucleus. The spectral line 
fits here are very suspect, as there is no sampling of the continuum to fit 
the line. In this situation, the line fitting algorithm systematically 
overestimates the continuum and underestimates both the line strength and 
velocity dispersion. As a result, estimates of the relative \ha\ flux and 
velocity dispersion in the emission line gas are only accurate outside the 
active nuclear region. 

Figures 1 (Plate 1) and 2 show the reduced \fp\ maps
for IR14348. Due to the low flux levels
in the off-\ha images, the continuum map shows little more than the two 
nuclei, separated by $\sim$ 4\arcsec. The \ha intensity map is much more 
complicated, with extended, clumpy emission throughout the body of the
system and in the tidal tails. Interestingly, the strongest \ha in the
NE member is found offset from the NE nucleus by 2\arcsec\ away from
the interacting companion. This offset is clearly visible in the both reduced 
maps and the raw images, and is not an artifact of the spectral fitting 
process. It is unclear whether this non-nuclear emission is intrinsically 
stronger than the nuclear emission or if the nuclear region is simply more
obscured -- more accurate photometry and reddening corrections are
needed to resolve this issue. Regardless, it is clear that the NE member is 
experiencing significant non-nuclear star formation.

The \ha emission is not confined to the main body of the system, either.
Several discrete clumps of \ha are found at large radius, $\sim$ 
5--15\arcsec\ (7--20 kpc) from the center of the system. These include a bright
emission source 5\arcsec\ to the south, extended emission in the NE
tail, and a system of four compact \ha knots in the SW tail. Using the
published \ha\ nuclear fluxes from Veilleux \etal (1995) to obtain approximate
photometric zeropoints, we can get an estimate of the \ha luminosity of 
these objects. The compact objects in the SW tail have \ha luminosities of 
$\sim 5 \times 10^{39}$ erg s$^{-1}$ each, while the more luminous object to 
the south has an \ha luminosity of $\sim 2 \times 10^{40}$ erg s$^{-1}$.
While the errorbars are large (factors of several), these \ha luminosities are
similar to those of the SMC and LMC respectively (Kennicutt \etal 1995).
The velocity dispersion and continuum levels in these regions are low, 
suggesting this is extended gas which has been compressed during the
collision and is now collapsing and forming stars. These regions may
represent the early stages of the formation of tidal dwarf galaxies
(\eg Zwicky 1956; Mirabel, Dottori, \& Lutz 1992; Hibbard \etal 1993).

From the derived \ha velocities, we measure a velocity difference between the 
nuclei of $\Delta v=125\pm 25$ \kms, consistent with previous measurements,
with the NE galaxy having a relative 
blueshift. The velocity map shows that the merging systems each still 
possess coherent rotation. Extracted one dimensional velocity cuts are
shown in Figure 3. The peak-to-peak velocity width of the NE galaxy is 
$\Delta v \sin i \sim 200$ \kms, while that for the SW galaxy is
smaller $\Delta v \sin i \sim 150$ \kms, probably due to a lower
inclination (recall that the SW nucleus is actually brighter in K by
0.54 mag; Carico \etal 1990). The extracted velocity cuts both run through
the bright \ha knot to the south, and it is here that the velocity
field shows a large ($\sim$ 100 \kms) shift to higher velocities.
How this region is dynamically linked to either or both of the galaxies
in the pair is unclear, but in both velocity cuts the higher velocity
\ha emission occurs after the rotation curves have flattened, suggesting
it is a distinct dynamical feature and not merely an extension of the
rotation curve(s).

The overall two dimensional velocity field indicates that, for the NE galaxy,
the encounter is fairly prograde. In both galaxies, the far side of the disks 
have a radial (line-of-sight) velocity in the opposite sense from the systemic
velocity of their respective companion, a classic signature of a prograde
encounter. More striking evidence
comes from the velocity structure of the (largely edge-on) NE tail. The
short, stubby tidal features produced in retrograde encounters typically
have large velocity gradients, as the material is stripped out of the disk
with a velocity in the opposite sense of the rotational velocity. The velocity
gradient in the NW tail (Figures 2 and 3) is quite small, inconsistent with
a retrograde origin. The fact that the tails are rather short ($\sim$ 20 kpc)
argues either that the collision is not {\it highly} prograde, that the
encounter is fairly young, or that projection effects are severe.

\subsection{IRAS 19254-7245 (The Superantennae)}

IRAS 19254-7245 (IR19254) is composed of a pair of galaxies, separated by 
10 kpc, and possesses an enormous pair of tidal tails which span 350 kpc 
from tip to tip (Mirabel, Lutz, \& Maza 1991).  The southern galaxy 
is a Seyfert 2 nucleus, with complex emission line velocity structure nearly 
1500 \kms\ in width (Mirabel \etal 1991; Colina, Lipari, \& Macchetto 
1991). Because the nuclear emission line complex is broader in width than
the spectral regions scanned by the \fp, no fitting could be done in 
regions in and immediately surrounding the nucleus; these regions have been
masked out in the final maps. Throughout the rest of the system, single
Voigt profile fits proved adequate to describe the observed line profiles.
The reduced \fp\ maps for IR19254 are shown in Figures 1 and 4.

Unlike IR14348, the strongest \ha emission in IR19254 is located within
a few kpc of the nuclear regions of the galaxies. The exceptions are
several \ha knots along the tidal arm of the northern galaxy and a single 
bright \ion{H}{2} complex on the south side of the southern disk.
In fact, the \ha flux from this bright \ion{H}{2} region is comparable to 
the \ha flux from the entire northern nucleus. From longslit spectroscopy, the
\ha luminosity of the northern nucleus is $10^{-14}$ erg s$^{-1}$ cm$^{-2}$
(Colina 1997, private communication), yielding an \ha luminosity for
the detached star forming knot of $L_{H\alpha} \sim 7 \times 10^{40}$
erg s$^{-1}$, a few times larger than that of the LMC (Kennicutt \etal 1995).  
HST imaging of IR19254 (Borne \etal 1997) shows a bright unresolved point 
source at this location; the inferred size is \ltsima 200 pc. The size
and luminosity of this object is similar to high surface brightness blue
compact dwarf galaxies (\eg Marlowe 1997). 

Because the disks in IR19254 appear more distorted than in IR14348, rather
than defining a major axis and extracting rotation curves, in Figure 5 we 
simply plot the velocities collapsed along the east-west direction -- a
position-velocity plot. Even with the morphological distortion, we see that
the velocity field of the system is fairly regular, showing two rotating
disks with velocity widths of $\Delta v \sin i =$ 400 \kms\ and 600 \kms\
for the northern and southern galaxies, respectively. Outside the central
regions the velocity dispersions are moderately low except for
the region between the nuclei. While the dispersion is large in this
region, there are clearly two components to the line profile (Figure 6a), 
suggesting that rather than true random motions, we are observing two
disks overlapping in projection. Both disks are rotating in the same
sense, with their northern sides receding. The northern galaxy is
redshifted compared to its southern companion (by $\sim$ 200 \kms),
confirming the expectation of a prograde encounter based on the presence
of the immense, linear tidal tails.

Extending to the east from the brighter southern nucleus is a plume of
\ha which does not correspond closely to any continuum features such
as spiral arms or bright knots observed in HST imaging of IR19254 (Borne
\etal 1991). The \ha line profile of this region
shows a strongly asymmetric blue wing (Figure 6b), suggestive of outflowing 
material. The luminosity weighted mean velocity of this plume is 
$v=17780$ \kms, or $v-v_{nuc}=-120$ \kms, where the nuclear velocity
is defined by the NaD absorption feature\footnote{Because of the multicomponent
emission line profile of the southern nucleus, defining an emission line 
redshift for this nucleus is problematic.} (Colina \etal 1991).
This inferred expansion velocity is much less than that of material in the 
nuclear regions, which have a velocity width of $v-v_{nuc}=\pm 800$ \kms 
(Colina \etal 1991). At a projected distance from the nucleus of
5\arcsec\ (6 kpc), the material in this extended plume must be slowing as it
expands outwards, as expected in starburst wind models (\eg Koo \& McKee 1992;
Heckman \etal 1996).
To get a simple estimate of the expansion time of this material, we note 
that for a spherically symmetric expanding superbubble powered by a 
constant mechanical luminosity, $r/v = (7/410)t_7$ where $r$ is the radius 
in kpc, $v$ the expansion velocity in \kms, and $t_7$ the expansion time in 
$10^7$ years (\eg Heckman \etal 1996). If the plume is expanding outwards at 
an angle $i$ from the line of sight, we have 
$t_7 = (410/7)(r_{obs}/v_{obs})\cot i$. 
Our derived expansion age for this \ha plume is thus $3\times 10^7 \cot i$ 
years, setting a rough timescale for the starburst activity.

Of considerable interest are the velocities of the faint knots at
large radius in the tidal tails, as they reveal whether the
knots are physically associated with the tails and, if so, 
give some constraint on the expansion velocity of the tails.  Because 
of the limited field of view of the \fp\ on the 4-m, only the the 
three brightest knots in the northern tail (see Figure 1 of Mirabel \etal 1991)
were observable, and of these knots only the closest had detectable
\ha emission (Figure 6c). The radial (line of sight) velocity of this knot is 
18030 \kms, redshifted by 100 \kms\ from the systemic velocity of the nuclei.
This detection is clear evidence that the interaction has triggered
star formation in the tidal tails 60 kpc from the merging pair.
Using the observed quantities, we get an expansion age of $r_{obs}/v_{obs} =
6 \times 10^8$ years, but this again is very crude, subject to projection
effects and geometry. However, a lower limit on the expansion age
can be estimated from the circular velocity of the disks; dynamical
models of merging galaxies show that the tail expansion velocity is not
larger than the circular velocity (Hibbard \& Mihos 1995; Mihos, Dubinski,
\& Hernquist 1997). Taking 300 \kms\ as an upper limit to the tail 
velocities, a lower limit to the expansion age is then $2\times 10^8$ 
years. It is clear that the starburst age (as measured by the \ha plume
expansion) is much shorter than the dynamical age of the merger (as
measured by the tails). Unlike IR14348, IR19254 ``waited'' a few dynamical
times after the initial collision before entering its ultraluminous phase.

\subsection{IRAS 20551-4250}

IRAS 20551-4250 (IR20551) is a single object with a relatively high surface 
brightness tidal tail to the south and a fainter tidal plume to the north.
The surface brightness profile is reasonably well fit by an $r^{1\over 4}$
law (Johansson 1991), except for in the central regions where it flattens; 
this profile is typical of the mass profiles of modeled merger remnants 
(\eg Barnes 1992; Hernquist 1992) where violent relaxation redistributes
the stellar mass in an $r^{1\over 4}$ mass profile. The fact that we see this 
quasi-elliptical light profile suggests that the violent relaxation
in IR20551 is largely complete and the two nuclei have assimilated into
one central mass.

The reduced \fp\ maps of IR20551 are shown in Figures 1 and 7. The continuum 
map shows the single body along with a bright ``ridge'' to the southeast -- 
probably a tidal arm or loop similar to those seen, for example, in NGC 7252 
(Schweizer 1982; Hibbard \etal 1995). The long tidal tail extending to the 
south is extremely diffuse, with none of the high surface brightness knots 
seen in IR19254 or IR14348.\footnote{Note that the bright compact object to 
the northwest is a foreground star (Johansson 1991).}
The \ha emission is concentrated in the central regions with only weak
extended emission visible. The notable exception is an \ha knot
to the southeast of the nucleus, where the tidal loop meets the main body.
Very faint emission is seen in the loop itself as well as in the tidal 
plume to the north; however, no \ion{H}{2} regions are seen
in the long southern tidal tail.

The \ha velocity field of the galaxy is very regular, and indicative
of simple circular motion, with a velocity width of 
$\Delta v \sin i = 150$ \kms. This rotating disk of ionized gas is similar
to those found in dynamical models of mergers, where gas which is 
not driven into the central regions settles into a warped, diffuse 
disk once violent relaxation has subsided (\eg Mihos \& Hernquist 1996).
The low velocity width for such a luminous system suggests we are 
viewing this disk largely face-on. If subsequent evolution is mild,
this disk may end up resembling the \ion{H}{1} disk found in the
nearby peculiar elliptical NGC 5128 (van Gorkom \etal 1990).
The only region that departs from the pattern of
circular motion is the region near the tidal loop where material stripped
out into tidal features is expected to be moving on very non-circular orbits
(\eg Hernquist \& Spergel 1992; Hibbard \& Mihos 1995).

Although no \ion{H}{2} regions are seen in the southern tidal tail, we 
searched for diffuse \ha by summing the pixel intensities in a 10\arcsec
x10\arcsec\ patch along the tail at a distance of 35\arcsec\ (30 kpc) from
the nucleus. Diffuse \ha is observed (Figure 8) at a velocity of 12500 \kms, 
blueshifted by 100 \kms\ from the systemic velocity of the remnant. At such
diffuse levels of \ha emission, it is unclear whether the ionizing source
is diffuse star formation in the tidal tail or the extragalactic UV background.

From a dynamical standpoint, IR20551 seems to be the most quiescent of the
ultraluminous IR galaxies we observed. The lack of a multiple nucleus,
the centrally concentrated \ha\llap, and the regular velocity field all argue
that the period of violent relaxation associated with the merging is
largely over, and that subsequent dynamical evolution will be slow.
Population synthesis modeling of the spectral features in the nuclear
regions by Johansson (1991) also suggests that the starburst event peaked
a few times $10^7$ years ago. While still ultraluminous, we may be catching 
IR20551 just past the peak of its luminous phase, as violent relaxation ends
and the starburst begins to fade. The coincidence between the end of violent
relaxation and the possible decline in the starburst intensity suggests
that ultraluminous starbursts do not continue long after the merger is
complete. 

\subsection{IRAS 23128-5919}

IRAS 23128-5919 (IR23128) is another double nucleus system, with a separation 
of 4\arcsec\ (3.5 kpc), and possesses a pair of tidal tails spanning 50\arcsec\
(45 kpc) from tip to tip. Bergval \& Johansson (1985) classify the nuclear
spectrum as intermediate between Seyfert 2 and LINER, and claim to detect
large scale outflow signatures in the form of asymmetric emission lines 
and an offset between the emission and absorption line velocities of the
nucleus. Johansson \& Bergvall (1988) also find Wolf-Rayet features in
the galaxy spectrum, indicating a strong on-going starburst. A bright
compact object is found near the end of each tidal tail, suggestive
again of star-forming knots; however, Bergval \& Johansson (1985) were
unable to detect emission from these knots.

IR23128 proved the most complex system for fitting
line profiles. Emission in the tails and the northern galaxy was well
fit by single Voigt profiles, but the line profiles in the southern
object were very complex, showing asymmetric shapes or double-line
profiles. In such cases, a single  symmetric Voigt profile is 
a poor model for the emission line, and these fits give poor values for the 
derived line parameters. For this reason, we chose to model obviously
distorted lines as a sum of two Voigt profiles. This choice is not meant to
imply two kinematic subsystems at each spatial point, but simply
to provide a better fit to the line shape (although in several regions
two distinct kinematic profiles {\it are} indicated by the data). With this
change in fitting procedure, the derived quantities take on a slightly
different meaning. The meaning of the continuum map is unchanged, while
the \ha\ intensity map is now the sum of the flux in the two fitted lines.
The velocity map now expresses the intensity-weighted mean velocity in
the line, and the dispersion map shows the second moment of the profile
shape. The reduced \fp\ maps for IR23128 are shown in Figures 1 and 9.

The continuum map shows the double nucleus and, faintly, the tidal tails
stretching north-south. Near the end of the tails, the compact objects
are easily visible. The southern galaxy is extended east-west; we argue
below that this structure is a rotating disk-like component, perhaps the 
original disk of the southern galaxy. In the \ha map, both nuclei show
strong emission, along with strong \ha between the two nuclei. Coincident
with the east-west extension of the continuum light, the \ha traces out a
flattened disk with strong spiral features around the southern nucleus.
Several \ion{H}{2} regions are found in the tidal tails. Interestingly,
the strongest \ha peaks in the tails do not correspond to the compact
continuum sources. No emission is detected from the continuum source in
the northern tail, while the southern object sits on the secondary \ha
peak in that tail. The strongest \ha peaks in the tails do not have detectable 
continuum emission. This lack of a strong correlation between continuum
peaks and \ha peaks suggests that the formation of tidal dwarfs is a 
stochastic process; continuum knots may be more evolved versions of \ha
knots, so that at any given time a spread in population ages is observed.
If so, this may make it difficult to finely date the ages of merger remnants
through the stellar populations in tidal dwarfs and globular clusters
(\eg Zepf \etal 1995; Schweizer \etal 1996).

The \ha velocity map reveals a rather chaotic velocity field; Figure 10 shows
the projected position-velocity plot. The tidal
tails are moving in opposite directions with a velocity spread of $\pm$100 \kms.
The velocity gradients along the tails are smooth, and indicate an axis of 
rotation along a position angle of 90\degns. In the southern galaxy we find
clear rotation, with a velocity spread of 400 \kms. The morphology and large 
velocity spread in this component suggests we are seeing it nearly edge on
and that it is rotating about an axis of position angle $\sim$ 0\degns. 
Rotation in the northern galaxy is less clear, although there is some
hint in the position-velocity plot that there is a weak retrograde sense
to its motion. The fact that this signature is weak, if present at all,
indicates that we see this system largely face-on.
In the region between the nuclei the velocity patterns
shift again; here the lines are broad and asymmetric, so it
is difficult to determine whether this is another shift in the net rotation 
axis or simply material decoupling from the overall rotation pattern due
to collisional shocks or winds.

The second moment (``dispersion'') map shows the largest velocity widths
occur along the edge of the rotating disk, away from the galactic nuclei. 
It is here that we see broad lines which often are broken up into two
distinct components (Figure 11). Much of the broadening of the line profiles 
must be due to the fact that we are viewing multiple kinematic components 
along the same line of sight. It is thus difficult to unambiguously determine
the presence of galactic wind in IR23128. The nuclear spectra on which
Bergvall \& Johansson (1985) based their wind detection had 6\AA\ resolution and
was taken through a 4\arcsec x2\arcsec\ aperture centered on the nuclei -- 
this aperture contains a significant amount of flux from the blueshifted portion
of the rotating disk, which may explain most, if not all, of the blue wing
seen in their low resolution spectra. To be sure, we also observe spectra 
with strong blue wings, but these are in regions near the disk component and 
may simply be due to line blending rather than true outflowing winds. Higher 
spectral resolution is necessary to resolve this issue.

The question of galactic winds aside, it is clear the dynamical state
of IR23128 is very complex. The axis of rotation shifts by 90\deg between
the outer tidal features and the rotating southern disk, while the northern
galaxy shows a somewhat retrograde motions, around a rotation axis orthogonal
to that of the southern disk. The tidal features generally trace much
of the initial angular momentum of the system; the fact that the inner
disk has been slewed by 90\deg suggests that the gravitational torques
have been strong and slow, \ie that IR23128 is well past the first encounter
and now experiencing the final destructive merging phase. The multiple
kinematic components within the merging pair cannot survive for long --
collisional shocks and gravitational torques will rapidly erase these features 
and drive strong inflow (\eg Mihos \& Hernquist 1996). We are most likely 
witnessing IR23128 in a very shortlived phase where violent relaxation is 
rapidly altering the structural and dynamical properties of the system.

\section{Discussion}

Although our sample of ultraluminous IRAS galaxies is necessarily small,
we nevertheless have detected a wide variety of dynamical states, even
though the bolometric luminosities are all quite similar.
The combined morphological and kinematic features in the galaxies suggests 
the following age-ordering for the four systems:

\begin{itemize}

\item   IR14348 is the dynamically youngest as indicated by its short
tidal tails, relatively unperturbed kinematics, and widely distributed
star formation.   The system has not merged yet and its first dynamical
response to the initial encounter is the initiation of wide spread star 
formation, which likely powers much of the observed IR luminosity.
Strong gaseous inflows may not yet have developed in IR14348, and this
system may be analogous to other ``young'' ultraluminous systems such
as II Zw 96, VV 114, and Arp 299 (\eg Yun, Scoville, \& Knop 1994; 
Goldader \etal 1997).

\item   IR19254 is well past the first collision as evidenced by
the long tidal tails.  The two disks are still well-separated 
($\sim$ 10 kpc) and their  coherent rotation  suggest it is not yet
in the latest stages of merging.  It is unclear if the Seyfert activity
in the southern galaxy is a response to the current interaction or
was a pre-existing condition.  The strong \ha\ seen near the nuclear
regions suggests the interaction is sufficiently advanced that gas is
starting to flow inwards.

\item IR23128 is dynamically complex.  It is currently experiencing
strong dynamical perturbation.  This coupled with  its close projected 
separation and distorted kinematics indicate it is the final merging phase
and most of the emission is concentrated in the central regions.  
Some evidence for kinematic feedback and baryonic blow out of the nuclear
gas exists and this system may therefore be in a similar state as
Arp 220 (\eg Heckman etal 1996).

\item IR20551 has a single nucleus and a quiescent velocity field.
These are the conditions expected in the post-merger phase of the 
interaction where only its remnant tidal features attest to a merger 
origin for the system.  Star formation is no longer widespread in the
system and its likely that its ultraluminous phase is near the end.

\end{itemize}

Given this putative dynamical age dating sequence, it is somewhat
surprising that at similar bolometric luminosity we see such a range
of dynamical conditions. Our sample was chosen to have high infrared 
luminosity, and one might have expected that this criteria would
have strongly biased our sample towards late stage mergers. While the 
sample is small, we do not see any strong bias. We do see a trend
between dynamical age and \ha concentration, in that more evolved
systems seem to have more centrally concentrated \ha emission (due to
central starbursts or AGN). This is consistent with the idea that collisions 
drive disk gas into the nucleus over a few dynamical times during the encounter.
Nonetheless, it is somewhat surprising that such high luminosities can
be achieved in young objects with extended emission as well as the more
evolved, concentrated objects. Clearly, ultraluminous activity
is not confined solely to late stage mergers where the gaseous inflows are
complete. 

At face value, the fact that these ultraluminous infrared galaxies are found
in a variety of dynamical phases might indicate that the ultraluminous phase
is long-lived. If the observed luminosity is powered by star formation, this
would require starburst activity to last several dynamical timescales.  Such
a scenario is rather difficult to envision.  Spectral modeling of ultraluminous
starbursts by Goldader \etal (1997) indicate starburst lifetimes of $\sim 10^8$
years, much less than typical merging times of $\sim 5\times 10^8$ years.
Furthermore, for inferred star formation rates of 100 M$_{\sun}$ yr$^{-1}$
(\eg Leitherer \& Heckman 1995; Goldader \etal 1997), the gas depletion times
are $\sim 5 \times 10^7$ years, again much shorter than the merging timescale;
truncating the IMF below 1 M$_{\sun}$ increases the depletion times by only a
factor of two (Goldader \etal). It would thus seem unlikely that sustained
ultraluminous starburst activity is possible in these systems.  Perhaps
a more stochastic starburst model, where a series of mini-bursts with
a long ``duty cycle'' occurs over a few dynamical timescales, might
supply the observed luminosity, although such a scenario seems rather contrived.
 
An alternative would be to appeal to AGN activity as the primary source
of the luminosity. In this case the timescale mismatch is significantly
lessened as sufficient gas is available to sustain AGN activity over a few
dynamical timescales. However, the relatively long time spent by interacting
galaxies at apogalacticon would result in a large fraction of ultraluminous
IRAS galaxies being observed in wide pairs if the ultraluminous phase
does span the merging timescale.  This prediction is in the opposite sense
of the observed morphological trends (Murphy \etal 1996), where most
ultraluminous systems are found to be close mergers.

Given that a long-lived ultraluminous phase is inconsistent both with
observational data and physical considerations, we are driven by our data to
the more likely view that dynamical phase is not the single triggering
criteria for ultraluminous activity. A variety of other factors must therefore
contribute to the formation of ultraluminous infrared galaxies.
While the merger mechanism is the root cause of the observed activity,
important physical details likely determine the particular luminosity
evolution of a given system.   Some of these other factors include 1) the
orbital geometry
of the encounter (\eg Barnes \& Hernquist 1996), 2) the internal structure
of the merging galaxies (Mihos \& Hernquist 1994, 1996), or 3) the
total amount of gas in the pre-merger disks.   Our small data sample
does not permit definitive conclusions but an important clue
comes from the fact that we are witnessing ultraluminous activity
in two galaxies which have yet to experience the violent relaxation associated
with the merger (IR14348 and IR19254), arguing that internal processes play
an important role in determining the response of galaxies to the encounter.
Both of these encounters are clearly prograde, as evidenced
by the long tidal tails and velocity field in IR19254 or the velocity field
of IR14348. IR14348 is  probably just past the first collision and
significantly dynamically younger than the other three systems.
Dynamical models show that
the strongest inflows (and, presumably, starbursts) occur during the final
merging, and are concentrated in the nuclear regions of the system (Barnes \&
Hernquist 1991, 1996; Mihos \& Hernquist 1994, 1996). While IR19254, IR20551,
and IR23128 fit this general pattern, why did IR14348 become ultraluminous at
an earlier phase, with star formation over a much larger spatial extent?

The answer may lie in the ambient properties of the disks.  IR14348 is
has the highest molecular gas mass of any of the objects in our sample
with an inferred molecular mass $\sim 6\times 10^{10}$ M$_{\sun}$  --
two to three times that of the other objects (Mirabel \etal 1990; Sanders
\etal 1991).  With such a large reservoir of gas, the perturbation at first 
encounter may drive gas into collapse {\it locally}, increasing gas densities 
and boosting star formation throughout the system without needing to wait for 
global instabilities to drive gas inwards and fuel a central starburst. 
As the system evolves, central gas flows will fuel a central starburst
thus changing the relative contributions of nuclear vs extended star formation
to the total emitted energy. Indeed, it is
interesting to note that the star forming morphology of this system is
similar to the morphology of high redshift galaxies in the Hubble Deep Field
(\eg Abraham \etal 1996; van den Bergh \etal 1996; Giavalisco \etal 1996);
IR14348 may represent a low redshift analogue of gas-rich star forming
galaxies in the early universe.

In our sample of ultraluminous infrared galaxies, we also find several
examples of extremely luminous \ha knots in tidal tails. These knots
have \ha luminosities comparable to that of the Large and Small Magellenic
Clouds, and their relatively low velocity dispersions suggest they are
pockets of gas in isolated collapse; if they have sufficient mass to
resist baryonic blowout, they may well evolve into bona-fide dwarf galaxies
(Zwicky 1956; Duc \& Mirabel 1994; Barnes \& Hernquist 1992). These objects 
seem to be more common in the less evolved
systems IR14348 and IR19254; by later stages of the merging evolution they
may have depleted their gas and ceased star formation or, in instances where 
these clumps form close to their host galaxies, fallen back into the merger 
remnant.   At the very least, however, our data clearly shows that objects
like dwarf galaxies can form in the early stages of a merger encounter at
distances up to 50 kpc away from the main merging bodies.

In summary, the velocity fields of ultraluminous infrared galaxies prove
quite diverse. There does not appear to be a common dynamical feature among
all four objects, other than the simple fact that they are all strongly
interacting. The variety of ages and kinematic features suggest that
several different factors play a role into triggering ultraluminous
activity, including interaction geometry, galactic structure, and gas content.
While numerical models do a reasonable job of describing the evolution
of galaxy mergers in a statistical sense, one-to-one matching of individual 
systems will prove more problematic due to varying initial conditions for
different interacting pairs. 
From a galaxy evolution standpoint, our data reveal the strong and rapid
dynamical evolution associated with violent relaxation in the final merging 
phase. The two objects which have not yet experienced this final phase --
IR14348 and IR19254 -- are comprised of relatively simple rotating disks.
On the other hand, IR20551 is perhaps only a few$\times 10^7$ years past its
merger, yet it already has settled down into a quiescent kinematic phase. 
Only IR23128 shows the very disturbed kinematics expected as two galaxies
ultimately merge, and the lifetime for this phase must be short. Over the
long dynamical history of a merger event, it is only in the final phase
that strong dynamical evolution occurs on global scales.  Once the merger 
is complete and the 
starburst quickly fades, the remnant will evolve rather quiescently;
however, signatures of the merger remain behind -- diffuse gas disks,
tidal features, and, possibly, young dwarf galaxies -- which attest to this
violent phase of galaxy evolution.

\acknowledgements

We thank Ted Williams for observing support during the run, and Charles
Beauvais for helpful discussions during data reduction. JCM is supported 
by NASA through a Hubble Fellowship grant \#~HF-01074.01-94A awarded by the 
Space Telescope Science Institute, which is operated by the Association of 
University for Research in Astronomy, Inc., for NASA under contract 
NAS 5-26555.

\begin{table}
\caption{ULIRG Sample}
\begin{tabular}{cccccc}\hline
Galaxy & Redshift & Recession Velocity\tablenotemark{a} & Distance\tablenotemark{b} & d(1\arcsec)\tablenotemark{c} & L$_{IR}$\tablenotemark{d}\\
 & (z) & (\kms) & (Mpc) & (kpc) & ($10^{12} L_{\sun}$) \\ \hline
IRAS 14348-1447 & 0.082 & 23715 & 310 & 1.4 & 1.8 \\
IRAS 19254-7245 & 0.062 & 17910 & 236 & 1.1 & 1.1 \\
IRAS 20551-4250 & 0.043 & 12660 & 167 & 0.8 & 1.0 \\
IRAS 23128-5919 & 0.045 & 13100 & 173 & 0.8 & 0.9 \\ \hline
\end{tabular}
\tablenotetext{a}{$V_r = c \times \sqrt{ ((1+z)^2 -1)/((1+z)^2 +1) }$.}
\tablenotetext{b}{for $H_0=75$ \kms\ Mpc$^{-1}, q_0 = 1/2.$}
\tablenotetext{c}{physical scale of 1\arcsec.}
\tablenotetext{d}{L$_{IR}$=L(8-1000\micron), see Sanders \& Mirabel 1997).}
\end{table}

\begin{table}
\caption{Observational Details}
\begin{tabular}{cccc}\hline
Galaxy & Wavelength Range & Number of & Velocity Coverage \\ 
 & Observed (\AA) & Images & (\kms) \\ \hline
IRAS 14348-1447 & 7098 -- 7115 & 18 & 775 \\
IRAS 19254-7245 & 6954 -- 6984 & 23 & 1375 \\
IRAS 20551-4250 & 6838 -- 6850 & 11 & 550 \\
IRAS 23128-5919 & 6842 -- 6868 & 24 & 1200 \\ \hline
\end{tabular}
\end{table}

\clearpage

\begin{figure}
{\it Color figure available at http://burro.astr.cwru.edu/preprints/ulirgs.html}
\vspace{20pt}
\caption{Reduced \fp\ maps of all four ultraluminous systems
observed. Top left: IRAS 14348-1447; top right IRAS 19254-7245 (The
Superantennae); bottom left: IRAS 20551-4250; bottom right: IRAS 23128-5919.
Each dataset shows the continuum map, the \ha\ emission map, the velocity
map, and the velocity dispersion map.} 
\label{f1}
\end{figure}

\clearpage

\begin{figure}
\plotone{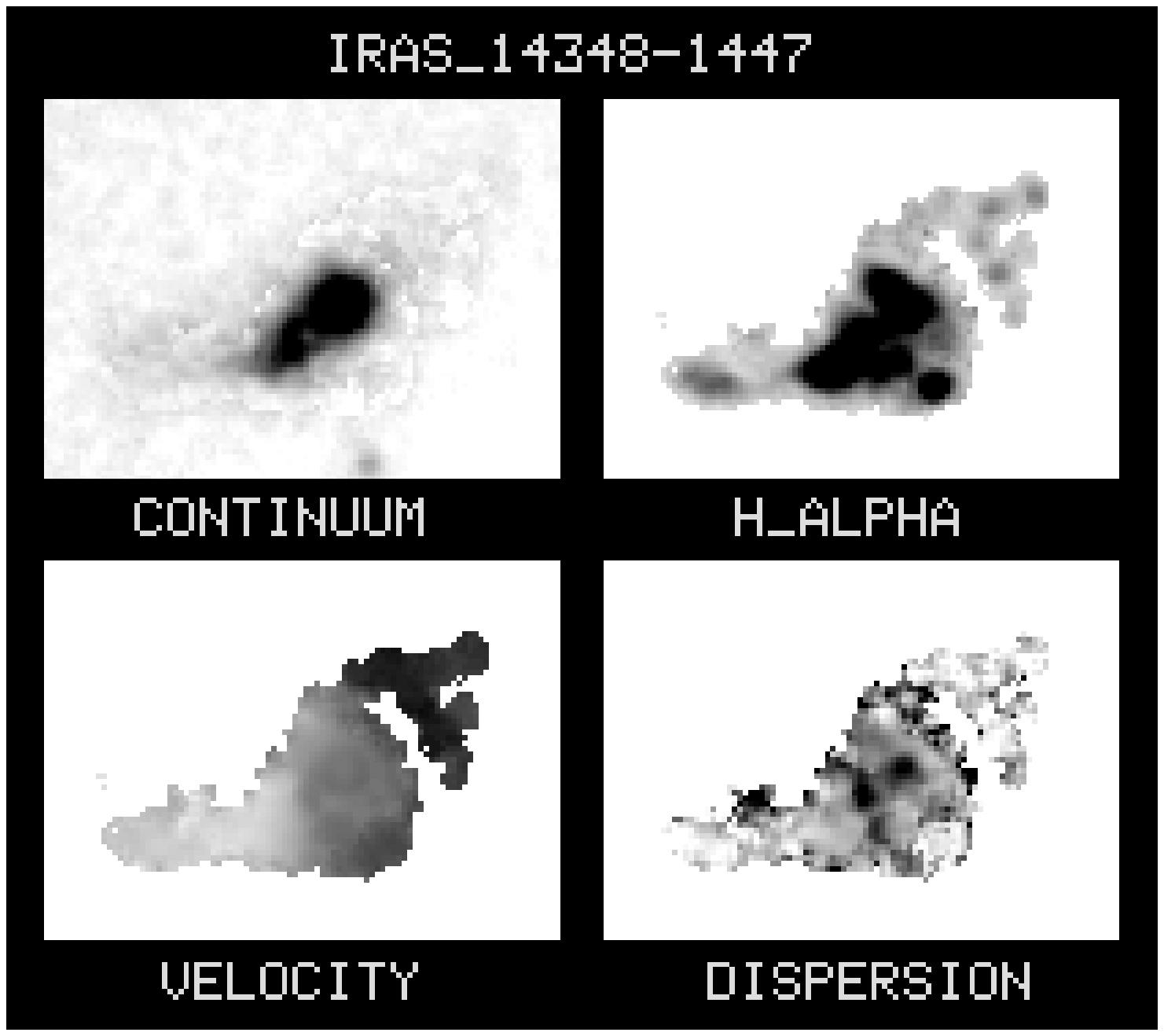}
\vspace{10pt}
\caption{Reduced \fp\ map of IRAS 14348-1447. North is to the left; east
is to the bottom. Top left: Continuum map.
Top right: \ha\ emission map. Bottom left: \ha\ velocity map. Light
shading is blue shifted relative to systemic; dark shading is redshifted.
Bottom right: \ha\ velocity dispersion map. Dark regions represent high
velocity dispersion.}
\label{f2}
\end{figure}

\begin{figure}
\plotone{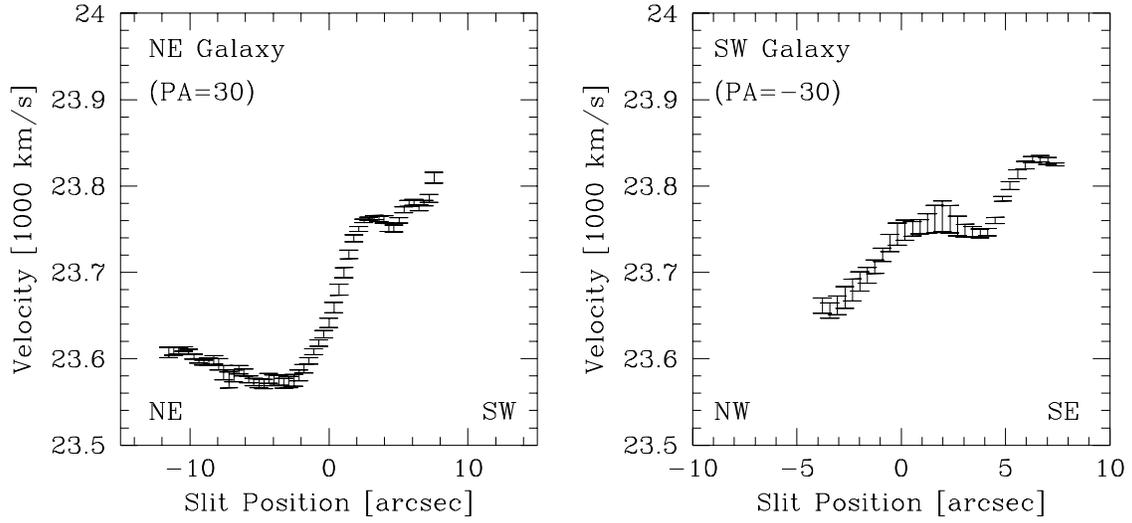}
\vspace{10pt}
\caption{Velocity cuts through the IRAS 14348-1447 system. Left: Velocity
cut through the NW galaxy at PA=30\degns. Right: Velocity cut through the
SW galaxy at PA=$-30$\degns.} 
\label{f3}
\end{figure}

\begin{figure}
\plotone{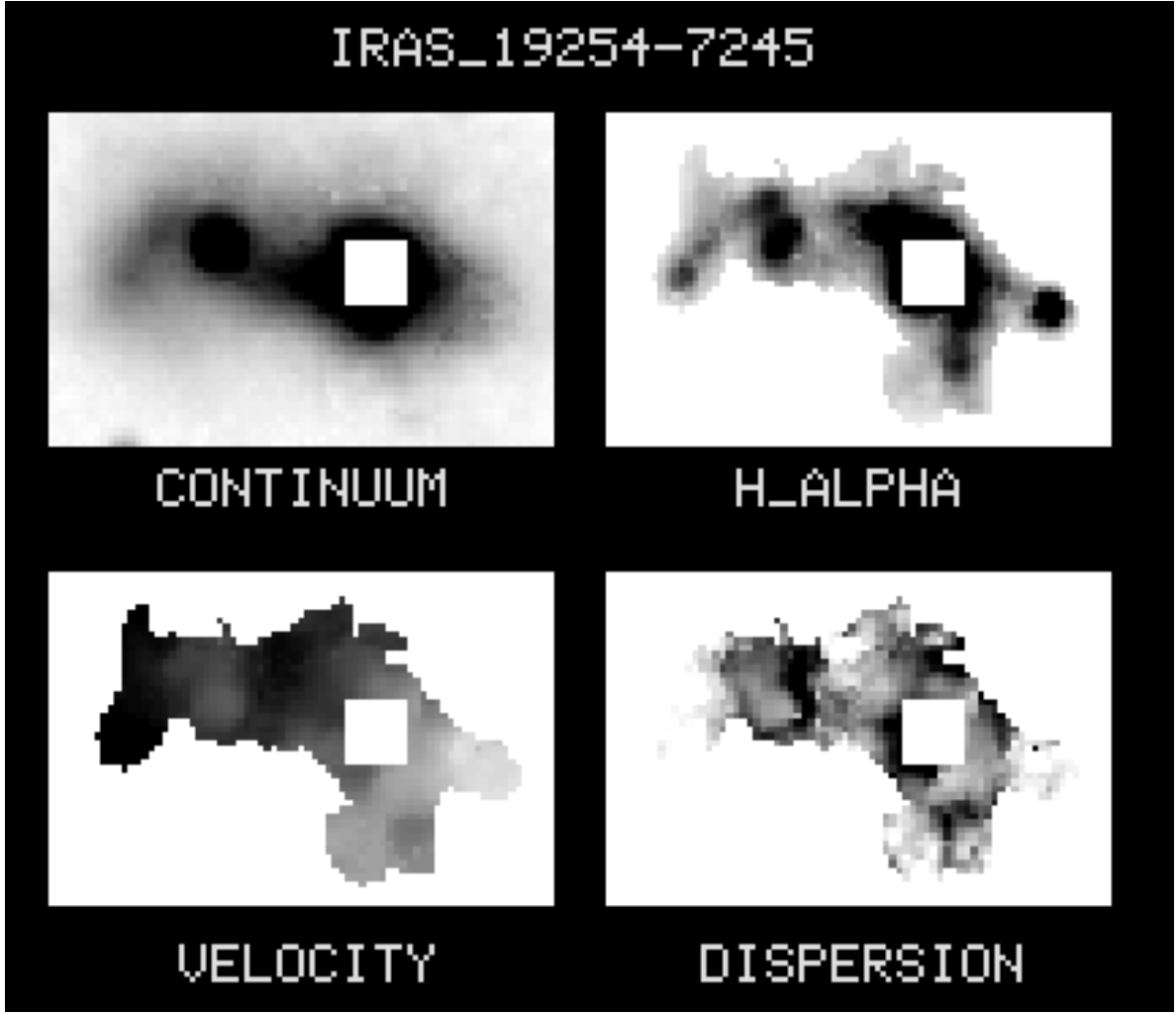}
\vspace{10pt}
\caption{Reduced \fp\ map of IRAS 19254-7245. North is to the left; east
is to the bottom. Top left: Continuum map.
Top right: \ha\ emission map. Bottom left: \ha\ velocity map. Light
shading is blue shifted relative to systemic; dark shading is redshifted.
Bottom right: \ha\ velocity dispersion map. Dark regions represent high
velocity dispersion. The central ``hole'' in the maps corresponds to the
Seyfert 2 nuclear region which could not be fitted by the line fitting
algorithm.}
\label{f4}
\end{figure}

\begin{figure}
\plotone{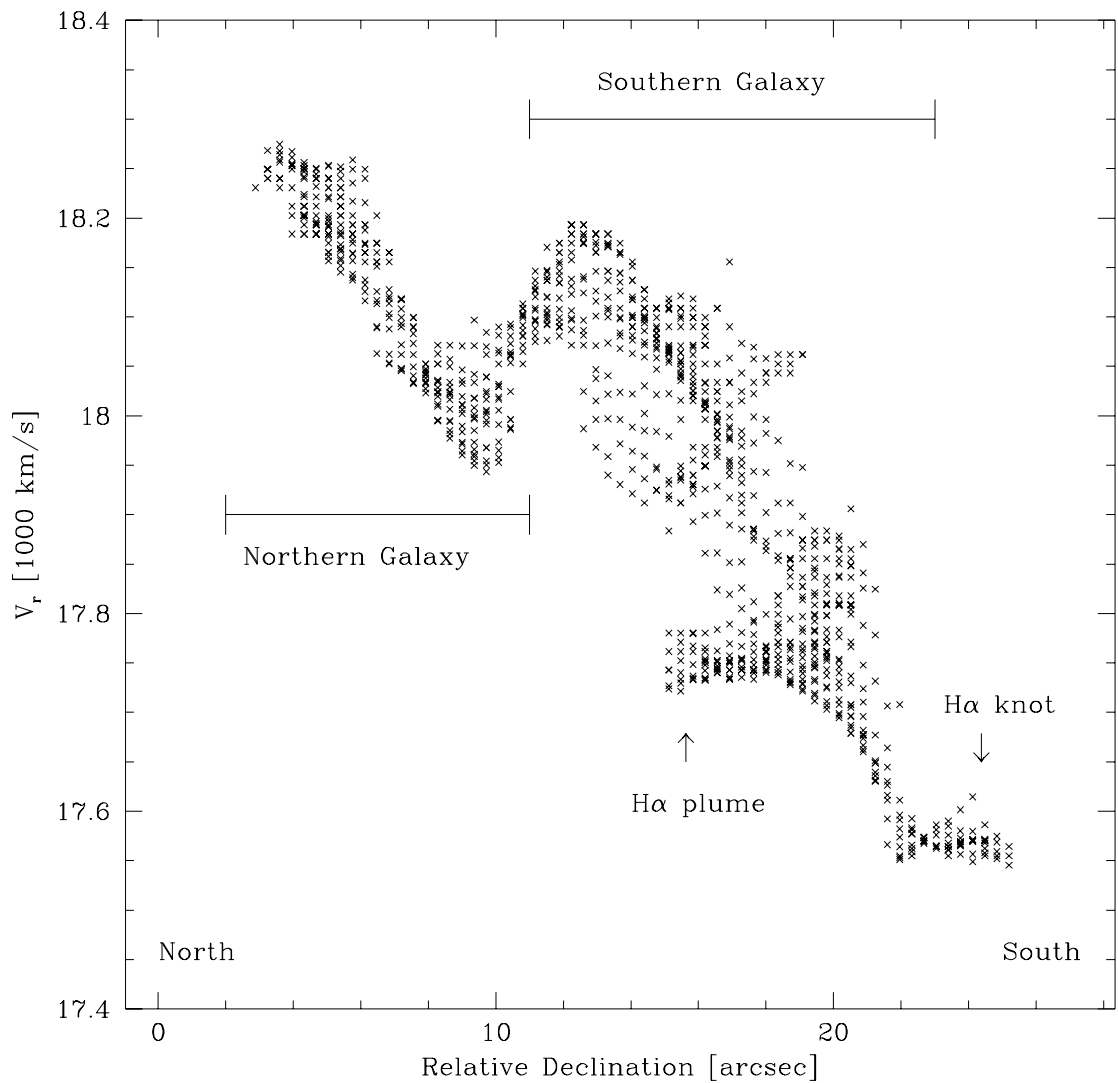}
\vspace{10pt}
\caption{Position-velocity plot for IRAS 19254-7245. In this plot, the 
2-dimensional velocity map has been collapsed in right ascension, yielding 
velocity as a function of declination. Two rotating disks can be seen,
along with a kinematically distinct \ha plume in the southern galaxy.
Note that where the two disks overlap, the velocities represent some
average velocity of the distinct disk kinematics (see Figure 6).
}
\label{f5}
\end{figure}

\begin{figure}
\plotone{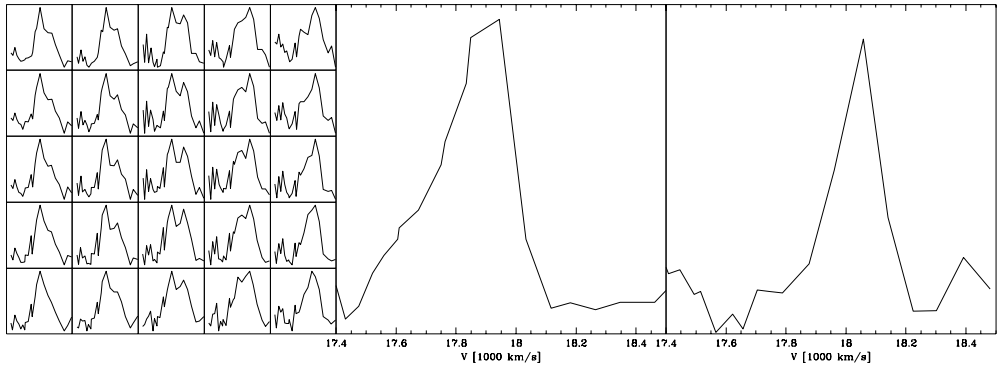}
\vspace{10pt}
\caption{\ha emission line profiles in selected regions of IR 19254-7245.
Left: The region between the two nuclei; notice the double line profile
indicative of two distinct kinematic features. Middle: The \ha plume
east of the southern Seyfert 2 nucleus. Right: The \ion{H}{2} region in
the northern tidal tail.}
\label{f6}
\end{figure}

\begin{figure}
\plotone{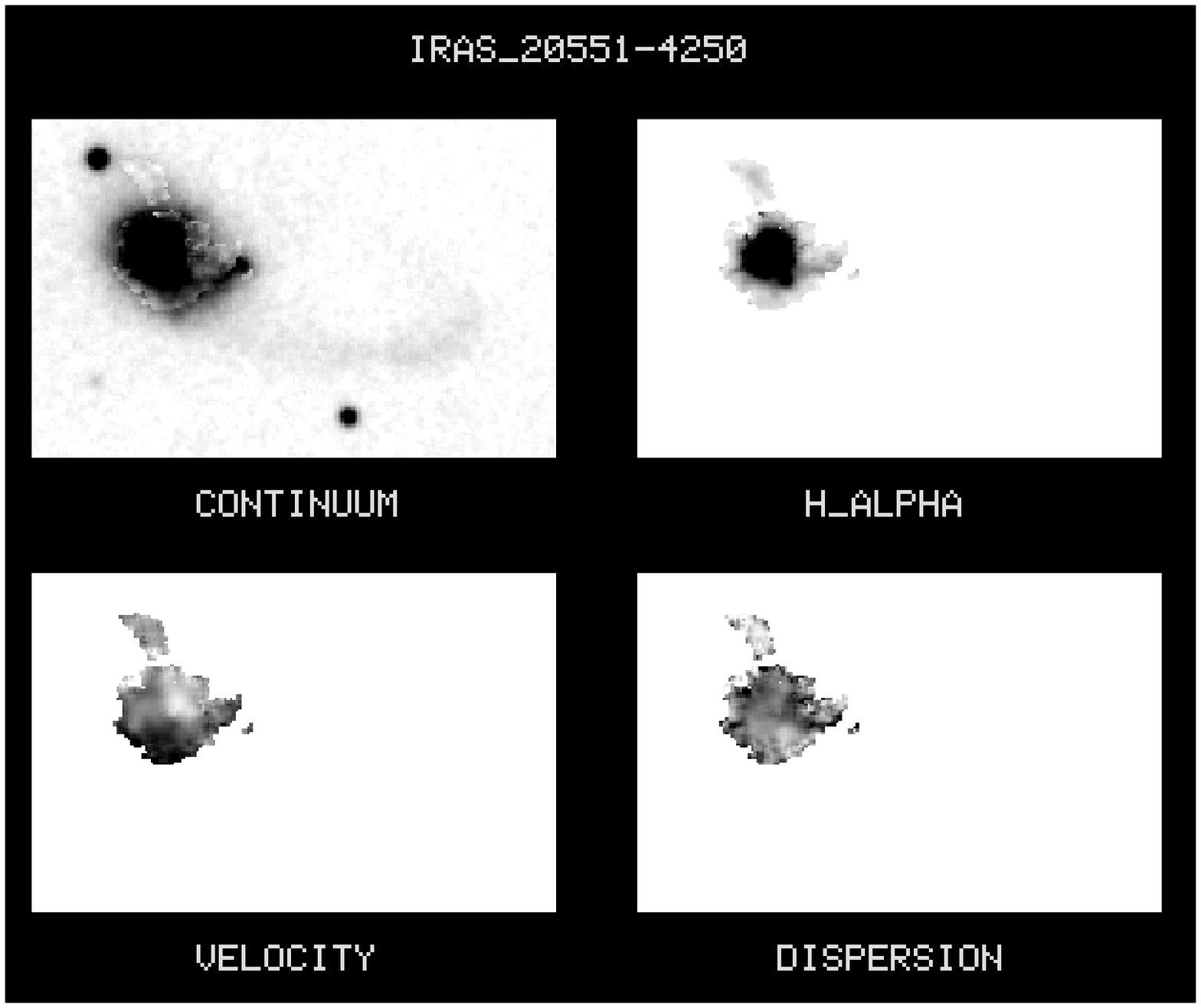}
\vspace{10pt}
\caption{Reduced \fp\ map of IRAS 20551-4250. North is to the left; east
is to the bottom. Top left: Continuum map.
Top right: \ha\ emission map. Bottom left: \ha\ velocity map. Light
shading is blue shifted relative to systemic; dark shading is redshifted.
Bottom right: \ha\ velocity dispersion map. Dark regions represent high
velocity dispersion.}
\label{f7}
\end{figure}

\begin{figure}
\plotone{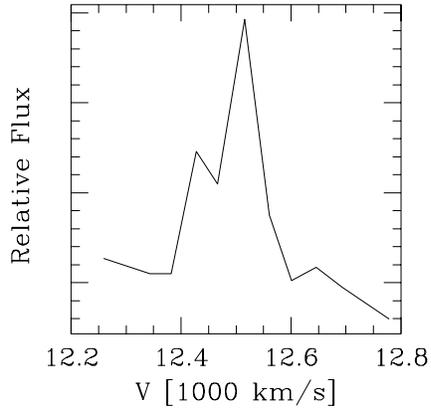}
\vspace{10pt}
\caption{The line profile of the diffuse \ha in the southern
tail of IRAS 20551-4250.}
\label{f8}
\end{figure}

\begin{figure}
\plotone{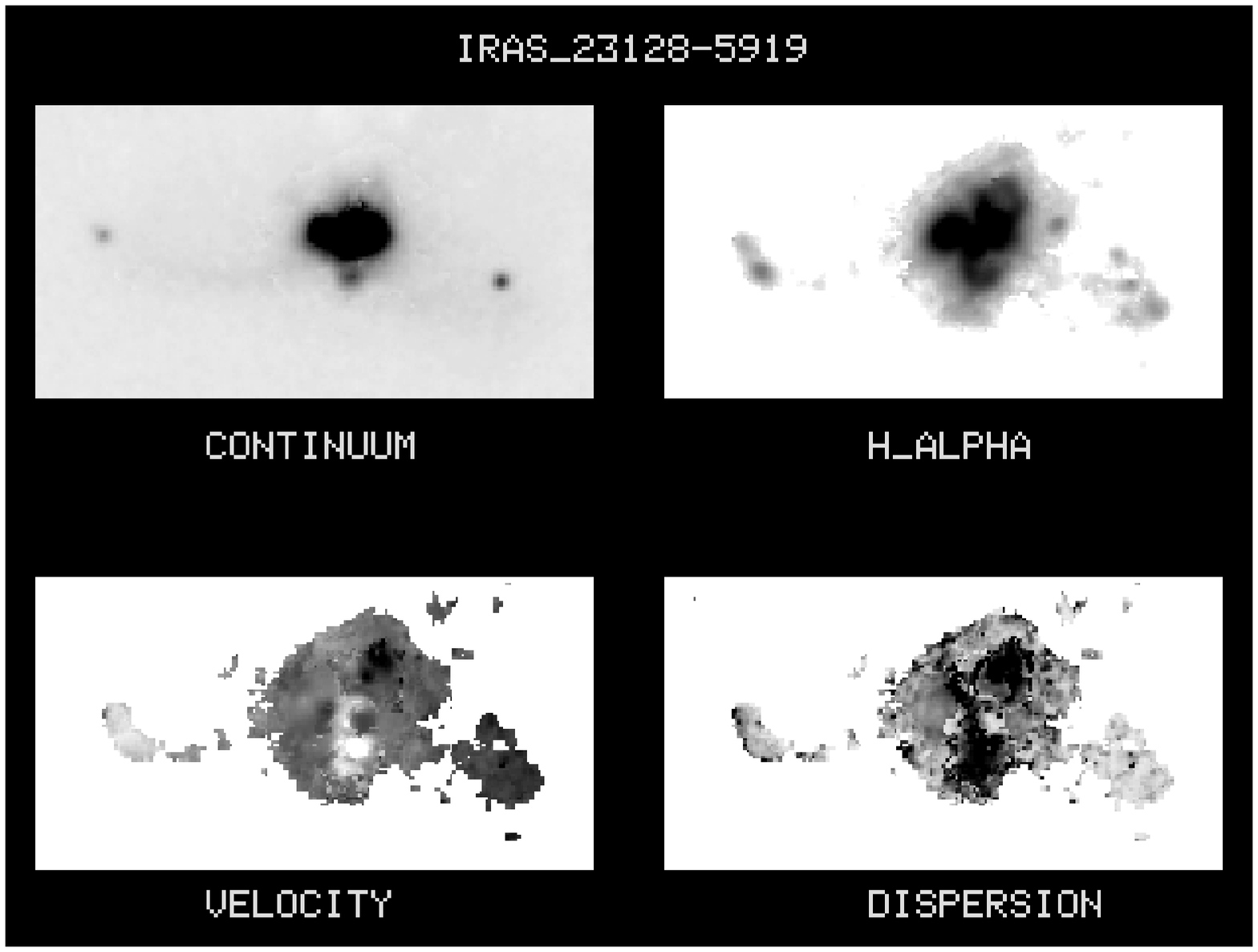}
\vspace{10pt}
\caption{Reduced \fp\ map of IRAS 23128-5919. North is to the left; east
is to the bottom. Top left: Continuum map.
Top right: \ha\ emission map. Bottom left: \ha\ velocity map. Light
shading is blue shifted relative to systemic; dark shading is redshifted.
Bottom right: \ha\ velocity dispersion map. Dark regions represent high
velocity dispersion.}
\label{f9}
\end{figure}

\begin{figure}
\plotone{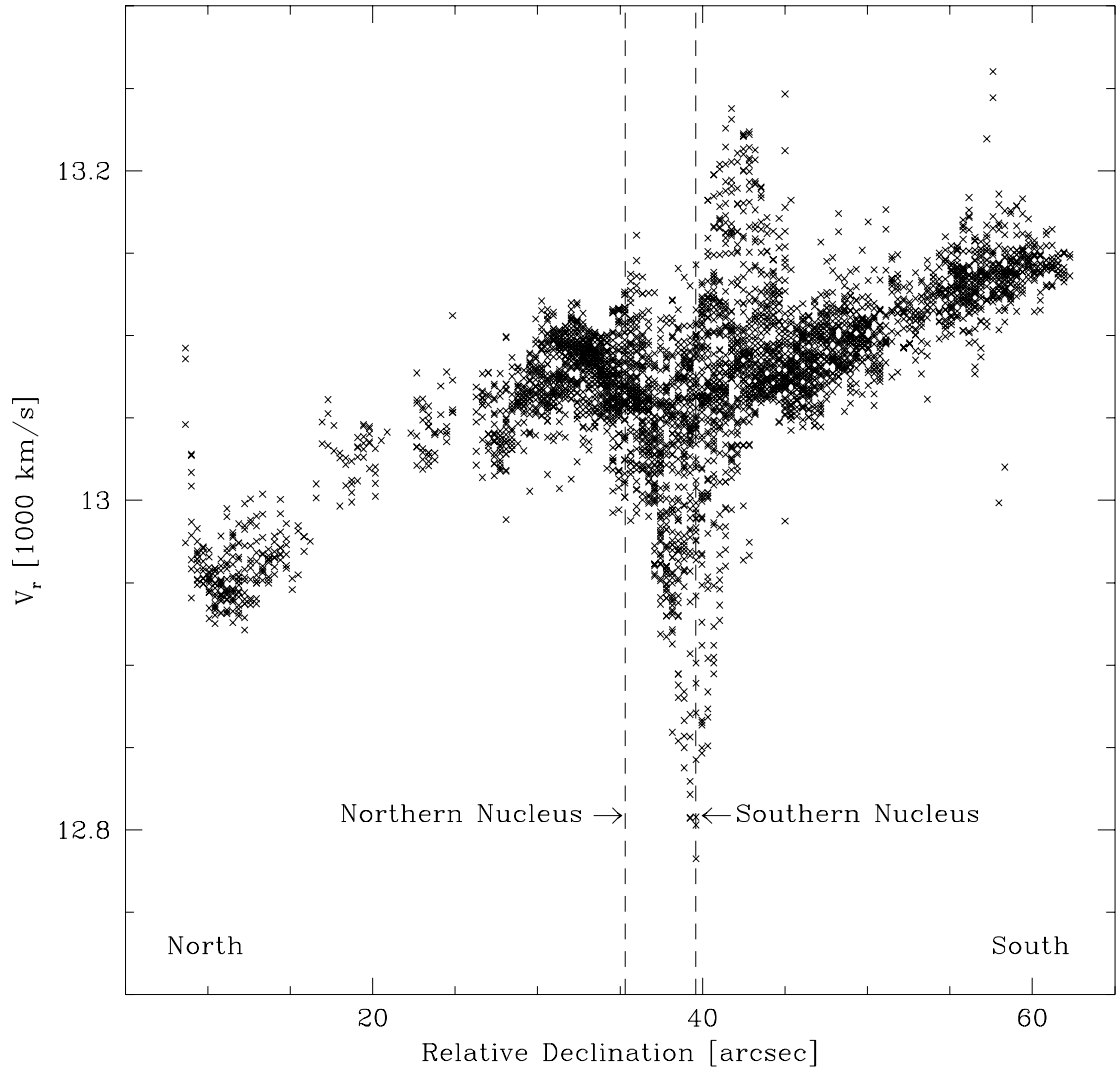}
\vspace{10pt}
\caption{Position-velocity plot for IRAS 23128-5919. In this plot, the 
2-dimensional velocity map has been collapsed in right ascension, yielding 
velocity as a function of declination. The dashed lines show the position of
the two nuclei. A rotating component is clearly
seen in the southern galaxy, and there is a weak signature of retrograde
motion in the northern galaxy. The velocity gradients in the tails are
also well-mapped.
}
\label{f10}
\end{figure}

\begin{figure}
\plotone{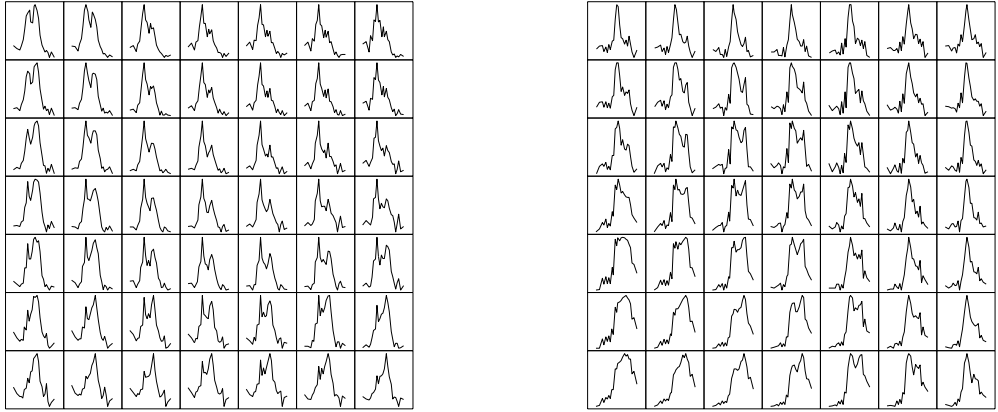}
\vspace{10pt}
\caption{Selected line profiles in IRAS 23128-5919. Left: Profiles in the
eastern, blueshifted portion of the southern disk. Right: Profiles in the
western, redshifted portion of the southern disk. Note the distinct kinematic
components in the spectra which shift in velocity.}
\label{f11}
\end{figure}

\end{document}